# BOUNDARY SPANNING AND THE SUPPORT OF DIGITAL ENTREPRENEURS: A CASE STUDY OF BAHRAIN

Noora H. Alghatam, University of Bahrain, nhalghatam@uob.edu.bh

**Abstract:** This paper explores the role of public and private sector teams as they collaborate to form and manage a community and platform for digital entrepreneurs in Bahrain. The paper employed the theoretical concept of boundary spanners to explore the nature of interactions between the two teams as support digital entrepreneurship and the outcomes that emerged from these interactions. The findings present the nature of the inter-sectoral interactions as boundary spanning that contributed to the initiation and formalization of the community and platform.

**Keywords:** digital entrepreneurs, public-private sector collaboration, ICT for development.

## 1. INTRODUCTION

Nations around the world have developed policies and institutional structures over the past few years to encourage and support entrepreneurship, and especially digital entrepreneurship. These national initiatives are driven by aspirations to leverage digital entrepreneurship capacities within local contexts in order for small businesses to scale and contribute to socio-economic development. Governments positively view the adoption of ICTs by entrepreneurs, since this is a means for businesses to innovate, scale in size and enter into new regional markets. As such, there are various initiatives to support digital entrepreneurship and many of them involve collaborations between the public and private sector. Such collaboration has led to the proliferation of accelerators, incubators, digital platforms and funding agencies in countries of the GCC region to support entrepreneurship and national development aims. The theme of collaboration processes between the public and private sector to support digital entrepreneurship is not researched enough, particularly from a sociotechnical perspective. This suggests the presence of a research gap that needs to be addressed. This paper explores the case study of Bahrain's entrepreneurial ecosystem, and particularly the setup of the "Startup Bahrain" community and a digital platform[1], which is a combination of a web presence and social media account, for the period of 2016-2019. The purpose of this study is to explore the boundary spanning activities of the public and private sector teams as they initiate, formalize and manage the community and platform, as well as the outcomes of this process.

This paper is structured as follows. First, we present a literature review on themes of digital entrepreneurship and development. We then outline the theoretical concept of boundary spanners. Third, we present the case study of the digital entrepreneurs' community and platform in Bahrain and the various institutional structures and actors that are involved. In the analysis of the case, we present the process of forming and managing the community for digital entrepreneurship that outlines the various boundary spanning actions taken by both the public and private sector. We then present the outcomes of boundary spanning and emergent outcomes.

## 2. LITERATURE REVIEW

---

[1] Our case study has some features of a platform, which is a web presence and social media account, that include some interactivity, data and complementary resources from the private sector.

## 2.1　Digital Entrepreneurship and Development

In recent years, many governments favourably view and support ICT adoption by entrepreneurs and small and medium sized businesses since this is considered as a means to improve economic growth (Boateng et al., 2008). E-entrepreneurship or digital entrepreneurship refers to people or entities that leverage entrepreneurial opportunities that are created from digital technologies (Davison and Vaast 2010). For the purpose of this study, we adopt the concept of digital entrepreneurship as both digital ventures that trade online as well as small and medium sized businesses that sell products and services through a combination of physical and online channels (Quinones *et. al* 2013). In recent years, there has been considerable interest in exploring the social embeddedness of digital entrepreneurs and how their actions and the socioeconomic context they are part of come to shape their activities and ventures (Avgerou and Li 2013). Some of these studies highlight the potential of ICTs to contribute to improve the performance of businesses and contribute to economic growth (Heeks 2002; Quinones, Nicholson and Heeks 2013; Chipidza and Leidner 2019). Moreover, these entrepreneurs and businesses that adopt ICTs are considered to potentially support in enhancing skills, generating job opportunities and improve innovation processes (Ramdani and Raja 2021). Even though some of the findings have shown diverse results from ICT adoption by SMEs (Rangaswamy and Nair, 2012; Quinones *et. al* 2013), there is still an overall acceptance of the potential for such projects to support socioeconomic development of countries (Chipidza and Leidner 2019; Sayed and Westrup 2003).

Academics discuss how digital entrepreneurship, scalability and digital platforms carry the potential to positively impact development (Avgerou & Li 2013; Neilsen 2017; Walsham 2017). In general, digital platforms are defined as sociotechnical ensembles of social and organizational processes and technical components (de Reuver *et. al* 2017). They can take the form of transaction platforms that act as marketplaces that facilitate information sharing and online transactions or innovation platforms that offer modular components that can be recombined to develop applications for the platform by third party entities (Cusomano 2019). In some contexts, governments support digital entrepreneurship through introducing platforms as a means to support the co-creation of services and encourage collaboration (Janssen and Estevez 2013). These platforms are often based on open data platforms that can be used to develop services within the public sector or for innovations by other sectors (Bonina and Eaton 2020). These platforms are used by citizens who require services and content, application developers and other government entities to provide content and get feedback from interaction by citizens and other groups (Janssen & Estevez 2013.) For example, Bonina *et. al*'s (2021) work on innovation platforms that support development and discuss the example of District Health Information Software (DHIS2) as a platform that offers data and software tools that could be used by developers to innovate solutions for platform-based services.

The public sector has also collaborated with various entities to support the entrepreneurial ecosystem as a whole. There have been calls to adopt a community perspective to explore digital entrepreneurship and to consider a broader concept that encompasses a collective of diverse set of actors with different aims and capacities (Nambisan, 2017). For instance, Du *et. al* 's (2018) works explores the case study of China adopts the lens of digital entrepreneurial ecosystem as a meta-theory to capture this group of actors involved as an entity. The interactions among actors within these communities or ecosystems are often discussed from a macro-level perspective in order to map out and measure their performance. Some academics also expressed the need to explore the local activities of actors involved in these ecosystems as they interact and engage in cross-boundary collaborations (Goswami *et. al*, 2018; Du *et. al* 2018). As such, we find that some studies explore how these actors, who come from different sectors, cross the boundaries of their professional domains and support the ecosystem for digital entrepreneurs (Feldman and Francis, 2006; Davison and Vaast, 2010; Nambisan 2017; Li *et. al*, 2017; Shen *et. al,* 2017, Goswami *et. al*, 2018). We consider the concept of boundary spanning as a useful conceptual lens to explore the interactions between members of the public and private sector team who work to support the digital entrepreneurs in Bahrain through the development of a community and digital platform.

## 2. THEORETICAL FRAMEWORK

We employ the concept of boundary spanners to explore the interactions of the public and private sector teams who come from different professional domains to collaborate and support digital entrepreneurship. Boundary spanners are individuals who span across boundaries, which can be professional domains, to communicate ideas and link groups with different locations and functions together (Tushman 1977). These spanning activities can link ideas, build relationships and set up fields between different groups.

These boundary spanners can have a bridging role, knowledge brokering role and can draw upon social capital as they engaged with other people across boundaries (Johri 2008; Milewski *et al.* 2008; Du & Pan 2013; Barner-Ramussen *et al.,* 2014). For instance, Johri's (2008) work employs the concept of boundary spanning as knowledge brokerage in the case study of engineers who work with others from different contexts. Other studies explain that boundary spanners are seen to be actors who come from different domains and draw upon cultural skills, social capital and language skills to have a bridging role between different disparate groups (Du & Pan 2013; Barner-Ramussen *et al.,* 2014). Milewski et al.'s (2008) also explores a case study of system engineers through the lens of boundary spanners who draw upon social capital and also play a bridging role as they work in two different countries.

The boundary spanning lens has been employed in the field of information systems to explore the processes of implementing digital entrepreneurial ecosystems (Milewski *et al.*, 2008; Li *et al.,* 2017). The concept is also relevant in Du et al.'s (2018) discussion of the case of entrepreneurship in China and how a platform offered resources and was a means for boundary-spanning practices. In this paper, theoretical the framework is employed to show nature of interactions between actors from the public and private sector teams as they work to support digital entrepreneurship. The aim is to explore how the team's engagement in boundary spanning contributes to the shaping of the community and platform for digital entrepreneurs (see figure 1). The paper addresses the following question: *How do the boundary spanning activities of the public and private sector teams shape the community and platform for digital entrepreneurs?*

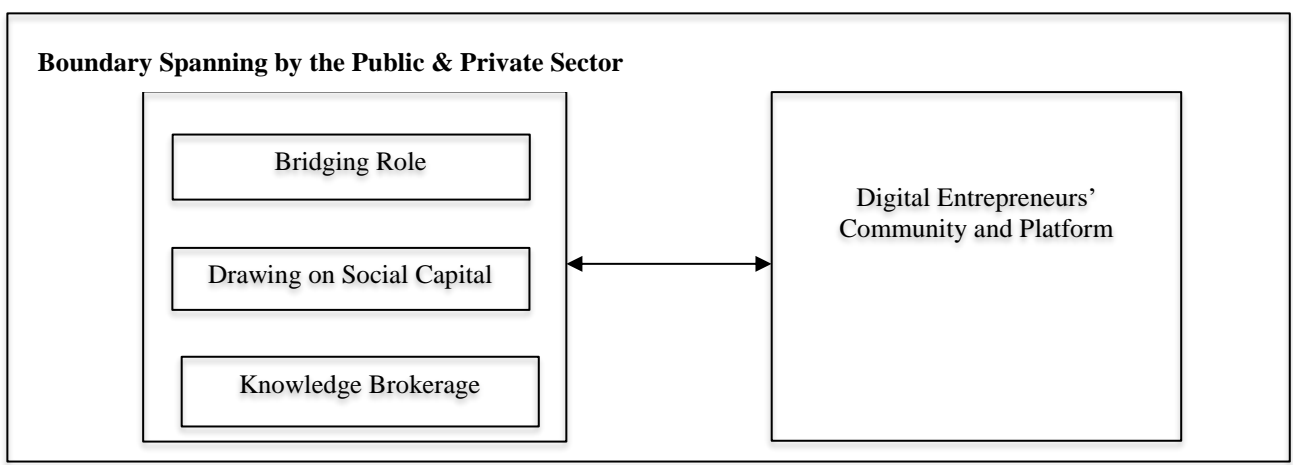

Figure 1: Boundary Spanning by the Public & Private Sector

## 3. METHODOLOGY

The empirical component for this research project was based on the national initiative to support entrepreneurship, and particularly ICT-based businesses, through the set up a community for start-ups. The research study began with a pilot study was conducted that focussed on the collection of secondary data and two interviews with a manager of an incubator and one of the academics from a think tank. The pilot study identified the key socio-cultural issues faced by digital entrepreneurs and also offered data about key milestones in the development of the community for digital entrepreneurship. The interview process included 17 semi-structured interviews that included managers and staff from both the public and private sector involved in the support of digital entrepreneurship. These interviewees were part of the different phases of developing the community for digital start-ups from planning to implementation and management. The data collection process also included four focus groups with digital entrepreneurs in Bahrain. Each focus group included 5-8 participants and were held at the University campus. The main criteria for the selection of participants in the study was that they were founders of a digital business and transacts online (offers online services) or a business that uses both physical and online channels to transact with customers.

The analysis of data was based on reviewing data from primary and secondary sources in a number of stages and iterations to identify key themes and milestones for the set up of the digital entrepreneurs' community. The data analysis was an iterative process as we compared concepts from primary data, secondary data and the literature on the topic. The early stages of analysis involved assigning general themes, or descriptive codes (Miles and Huberman 1994), to the data based on important activities of the public and private sector and significant milestones for digital entrepreneurs' community. The data analysis then shifted to focus on theoretical constructs of boundary spanning and linking them to the data. For example, there were concepts such as bridging roles, knowledge brokerage and drawing on social capital identified. Then as these themes were identified, pattern codes were developed such as "community building" and "information exchange" and "community outcomes".

## 4.  CASE STUDY

### 4.1  Institutional Structures for Digital Entrepreneurship

This section presents key milestones to support digital entrepreneurship in Bahrain for the period of a decade from 2008 to 2018. One of the most significant milestones was the development of the "Startup Bahrain" initiative in September 2017. The Startup Bahrain initiative represents a community that included a number of entrepreneurs, key actors from the public and private sector to network, organize events and meetings and support the entrepreneurial community (see figure 2). The Startup Bahrain community included the creation of a leadership team and the articulation of six pillars for the ecosystem to enhance and maintain the growth of the community. One of the most influential actors in the public sector was the Economic Development Board (EDB) which supported the development of this community (www.edb.com). The emphasis on digital entrepreneurs was in line with a study of high potential startups, which were businesses that had the potential to scale globally, especially through digital platforms. These high potential startups were in their early phases of operation and needed opportunities to network in order for them to generate ideas and obtain advice. The aim was to set up a community, and then over time, have the startups themselves take over the organization and management of the community. As the initiative progressed, there was a realization that having a leadership team would be useful to manage the community. The leadership team included members from the public sector, accelerators, incubators and educational institutions. The team's role was to identify the needs of the community since they were engaged in communication with them through events and this data would help design and improve the community for digital entrepreneurs and the overall entrepreneurial ecosystem. This came in line with the EDB's six pillars for the ecosystem which included: community, incubators, accelerators, funding, education, corporates, policy and regulation.

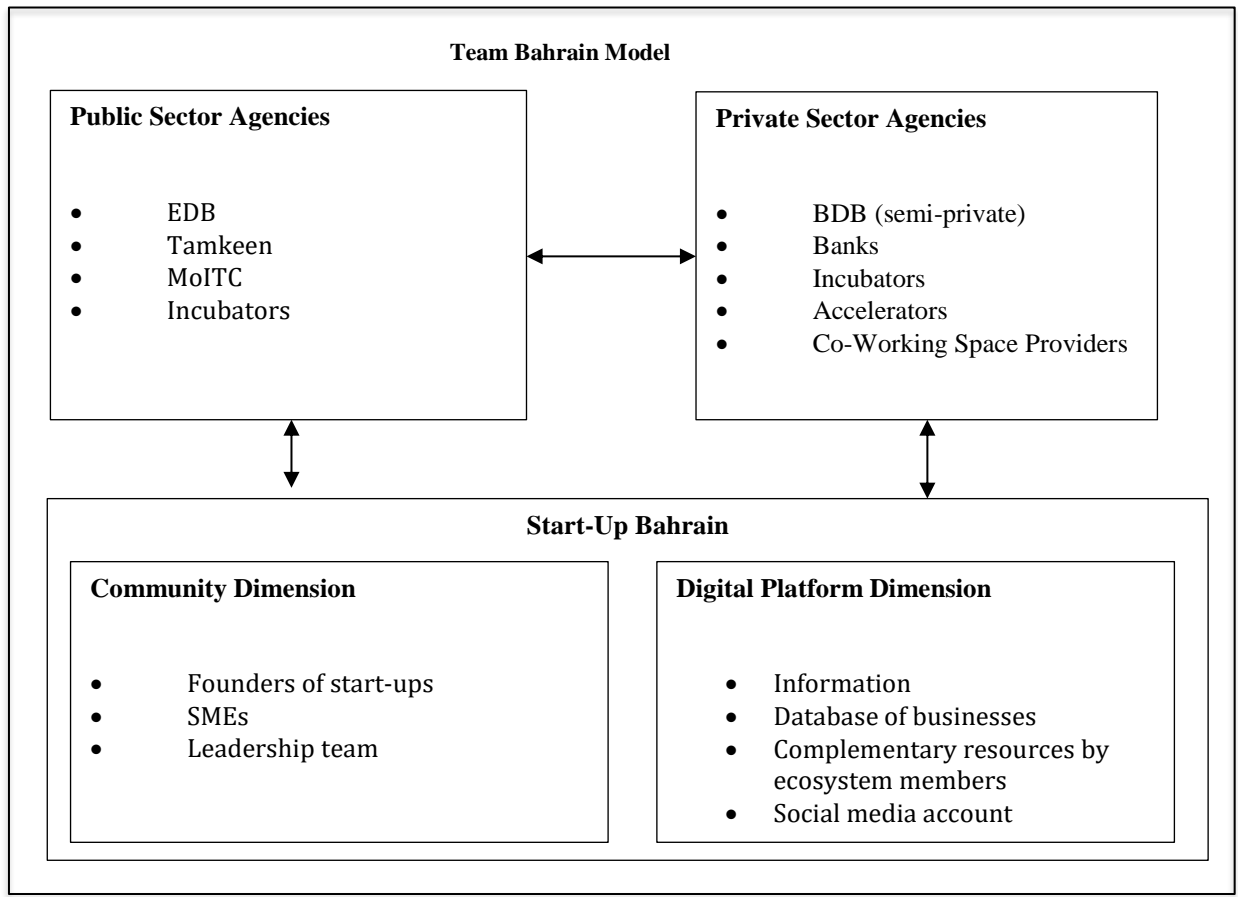

**Figure 2: Entrepreneurial Ecosystem**

### 4.2 Economic Vision and Regulations:

The efforts to set up a community for digital entrepreneurs is in line with an existing national policy, which is the Economic Vision 2030 (www.bahrain.bh). The policy includes many areas for economic development and there are two themes that are in parallel to digital entrepreneurship. One is the economic pillar that emphasizes the need to diversify income and high potential markets. Second, is the government pillar that includes concepts such as private sector partnerships and outsourcing. Over the years, there were various developments that positively affected digital entrepreneurship such as new regulations and institutional structures to offer funding (Lopez *et. al* 2020). The regulatory changes included the development of virtual commercial registration, Sijili, which supports the setup of commercial registration for businesses that do not have a physical office, i.e., businesses can be registered based on a virtual office. Another set of regulatory developments include the cloud first policy and the equity crowd funding regulation, which enables the operation of electronic lending platforms and supports Fin-tech startups. In subsequent months there were other regulations that were conducive to the entrepreneurial ecosystem such as competition law, bankruptcy law, the health insurance law and the data protection law. There were also a number of entities that offered financial support for entrepreneurs such as The Bahrain Development Bank (BDB), a semi-government bank that offers financing schemes to entrepreneurs. Another significant government entity is Tamkeen that offers a combination of financial and educational programs to support entrepreneurs.

### 4.3 Startup Bahrain as a Digital Platform:

A significant step in the formation of a community for digital entrepreneurs was the development of the Startup Bahrain digital platform that included information, a database and resources that served the entrepreneurs, investors and other stakeholders. The current digital platform was planned and implemented by the EDB with the support and involvement of some private sector agencies and entrepreneurs. The current platform includes information on investments in Bahrain to encourage foreign investments and international businesses to set up in the country. There is also information about latest laws and regulations that facilitate the deployment of startups and upcoming events hosted by the community. The website also includes a list of important actors in the entrepreneurial ecosystems such as government agencies, IT and management service providers, co-working space providers and funding agencies with links and contact details. Finally, the platform includes a database of digital entrepreneurs and the option of signing up your business to be included in the database.

Even though the digital platform Startup Bahrain was mainly run by a public sector agency, there are a few features that set it apart from existing e-government websites. First, as noted by interviewees, the objective of the community and associated platform was to have a transition from being public-sector run towards being independently managed by the community as the initiative gains momentum. Second, the platform aims to include the community's involvement and interactivity, and this is depicted in the current blog and social media account that includes live interviews with entrepreneurs and stakeholders. Such an emphasis on encouraging community involvement and interactivity marks a distinct difference with other government websites. Third, the digital platform showcases existing partnerships with the private sector to support entrepreneurship through offering resources that support them such as Amazon Web Services credit and resources that entrepreneurs could apply for. Other supportive resources are offered by financial and public relations agencies involved in the ecosystem.

## 5. ANALYSIS

### 5.1 Boundary Spanning to Initiate the Community and Digital Platform

Members of the public sector who were supporting entrepreneurs played a bridging role that brought together various stakeholders from the private and public sectors and this included entrepreneurs as well. As noted during an interview by a manager in the public sector, one of the initial steps taken was identifying the needs of digital entrepreneurs and defining the nature and objectives of the community. The public sector actors realized early on that they needed to focus on high potential startups, many of which were businesses that could leverage digital technologies and platform to scale globally. The establishment of the community included further actions. First, identifying gaps in the existing context for digital entrepreneurs. From this process they decided to establish a community with diverse members and to promote the community in both local and international context. Second, the public sector created a leadership team from various sectors. The public sector assessed the environmental gap that entrepreneurs faced and based on this they formed a leadership team to set objectives and promote the community in society and internationally. This leadership team included businesses and entrepreneurs who were actively participating in digital entrepreneurship early on, as well as businesses and public sector entities that support entrepreneurs through mentoring and funding. The central role of the public sector was evident as they hosted the digital platform, that was a combination of a web presence and social media account, which was developed for Startup Bahrain and included information about policies, upcoming events, blog posts, resources and ecosystem players.

Subsequently, the private sector team involved in the set up of the community were actively playing a knowledge brokerage role as they linked the community with international networks of experts and businesses to support the entrepreneurs. As noted in an interview with a manager in one of the accelerators, his company invited experts from other international branches of their business to give presentations and act as mentors in community events. The international experts often took part in community-focused conferences, workshops and mentorship events to transfer knowledge. The idea was that transferring knowledge on entrepreneurship would contribute to building the capacities of digital entrepreneurs. The aim was to support the community's members to scale their businesses and enter into new markets. The vision was that once they gained momentum, the entrepreneurs would take over the management of the community itself. This knowledge brokerage role was also strengthened further with the digital platform that included complementary resources from local and international companies. For instance, there were free credits and supportive material offered by local marketing agencies as well as international platforms such as Amazon web services. Through these resources, the platform reflected and reinforced the public and private partnerships even further.

The formation of the community also led to processes of drawing upon social capital by both the public and private sector to support the scaling of entrepreneurs in the community. Both sectors were drawing upon their understanding of sociocultural institutions such as they role of a business network for high-potential startups. Many of these small businesses were ICT-based and had founders who are fresh graduates or new to the market and needed a community that represents them. This theme was prevalent in discussions during focus groups where participants noted that they valued the social support. The aim was to establish a narrative of entrepreneurial success to ensure values of societal trust was established for the digital entrepreneurs. This was so the concept of an entrepreneur could be reframed in society from being conceptualized as a small local store to a digital business that could scale and contribute to economic growth. This boundary spanning activity of drawing upon social capital that informed the process of creating a narrative was reflected in the platform that included a database of startups, upcoming events and blog posts about entrepreneurs. Moreover, the social media account fostered interaction within the community and included live interview sessions with community members to showcase their businesses and experiences (see Table 1).

| Boundary Spanning | Actions | Outcomes |
|---|---|---|
| Bridging Role | Link stakeholders for community building<br>Identifying environmental gaps<br>Focus on high-potential start ups | Develop a formal community<br>Set up a platform with information and complementary resources.<br>Establish a leadership team |
| Draw on Social Capital | Identify the need for a supportive network.<br>Organize events<br>Discourses for success in events<br>Online Interviews | Development-focus values for entrepreneurs<br>Reframed entrepreneurship concepts<br>Central role of the public sector |
| Knowledge Brokerage | Link experts from other business branches<br>Invite experts as mentors and speakers<br>Offer complementary resources on platform. | Intermediary role between the local and global networks.<br>Digital platform as encompassing public and private partnerships through resources. |

**Table 1: Boundary Spanning Activities**

### 5.2   Outcomes for the Community and Digital Platform

The public and private sector teams' interactions came to shape the community for digital entrepreneurs in two ways. First, the boundary spanning activities of the public sector and drawing upon social capital such as policies and development goals led to the establishment of a developmental focus for the community. The public sector was keen to enrol high potential start ups, which were ICT-based businesses that could scale in size and enter into new regional markets. This was considered as a means to support socio-economic growth policies that aims to support Bahrain's position as an innovation hub in the region. As explained by entrepreneurs during focus groups, a motivational factor for developing a business was to contribute to national socioeconomic growth and job creation.

Second, the boundary spanning activity of bridging between various stakeholders to form the community and platform led to enforcing the central role of the public sector in entrepreneurial activities. From the outset, there was a clear objective to have a transitional role for the public sector until the community grows and takes over the management process. The digital entrepreneurs accepted the public sector's role since they were in need of a supportive network for ICT-based businesses and also a change in societal narratives to establish of trust of entrepreneurs. The public sector played a pivotal role in setting up events and programs to overcome such challenges to establish and support the community. The public sector also sees themselves as embedded in the community itself and this initiative was more than a short-term project. As noted during interviews, they were working on building a conducive environment for business. The digital platform was used to leverage intermediary roles between local and international experts on entrepreneurship. This was evident in the complementary resources offered on the platform and interviews and events, many of which are currently hosted online.

## 6.    DISCUSSION AND CONCLUSION

The study's findings highlight the boundary spanning activities of the public sector which bridged different groups, that included the private sector, other government agencies and start-ups, to initiate and formalize the Start-up Bahrain community and platform. Even though this was a collaborative initiative between various sectors, it was evident that the public sector plays a dynamic role in facilitating intersectoral collaboration to support the formation of the digital entrepreneurs' community. This contributes to existing studies that emphasize the central role of regulatory actors in national ICT initiatives (Du *et al*., 2018; Li *et al*., 2018). The process of boundary spanning evolved over time to reflect intersectoral collaboration that included emergent team capabilities of linking local and international networks of experts on entrepreneurship. The organizing of conferences, mentoring sessions and judging panels in events to include entrepreneurship experts from other contexts shows an emergent team capability. This contributes to existing studies of the role of local and global networks in the shaping of digital entrepreneurship (Avgerou and Li 2013; Quinones *et. al* 2020).

The boundary spanning activities came to contribute to emergent roles and values for the community. This contributes to studies that call for an emphasis of the social embeddedness of digital entrepreneurship (Avgerou & Li 2013). Indeed, the actions of the public and private sector teams were shaped by an understanding of regulations and economic policies, gaps in the entrepreneurial environment and societal discourses on entrepreneurship. As such their bridging and knowledge brokerage role came to support the creation of a network for these high potential start-ups and a reframed discourse of success for start-ups.

In terms of theoretical outcomes, this paper contributes to existing studies of ecosystems, institutional structures and networks that support digital entrepreneurs in order to attain socioeconomic development. The paper's contribution is adoption of a micro-level approach to explore the interaction between various actors to support the digital entrepreneurs. This comes in contrast to existing studies that adopt a macro-level perspective on factors and antecedents that support the development of a digital business environment. The paper also presents practical contributions that can be translated into suggestions for public and private sector practitioners involved in such initiatives. The public sector plays a role in transferring and supporting ICT4D values through interacting with the private sector players and the digital entrepreneurs. The findings suggest that the platform plays a role in supporting boundary spanning activities for cross-sectoral collaboration. One suggestion is to leverage the current platform to include open data sets and modular components that could be offered for entrepreneurs and stakeholders to use to develop further innovations and support their activities and the community as a whole.